\documentclass[sigconf,nonacm,natbib=false,letterpaper,final]{acmart}

\usepackage{glossaries}
\usepackage[commandnameprefix=always,todonotes={textsize=small,obeyFinal}]{changes}
\usepackage{siunitx}
\usepackage[capitalise]{cleveref}

\usepackage[backend=biber,maxbibnames=99,minbibnames=1,mincitenames=1,firstinits=true,sorting=none,style=ieee,doi=true,eprint=false,isbn=false,url=false,style=acmauthoryear]{biblatex}
 \usepackage{graphicx}
\newacronym{bss2}{\mbox{BSS-2}}{Brain\mbox{ScaleS-2}}
\newacronym{hpc}{\mbox{HPC}}{High-Performance Computing}
\newacronym{esd}{\mbox{ESD}}{EBRAINS Software Distribution}
\newacronym{ci}{\mbox{CI}}{Continuous Integration}
\newacronym{nv12}{\mbox{NV12}}{NVLink~12}
\newacronym{nv4}{\mbox{NV4}}{NVLink~4}
\newacronym{abi}{\mbox{ABI}}{Application Binary Interface}
\newacronym{e4s}{\mbox{E4S}}{Extreme-Scale Scientific Software Stack}

\AtBeginDocument{%
  }

\begin{document}

\title[HPC Containers for EBRAINS: Towards Portable Cross-Domain Software Environment]{HPC Containers for EBRAINS:\linebreak[2]{}Towards Portable Cross-Domain Software Environment}

\author{Krishna Kant Singh}
\orcid{0000-0002-7472-6299}
\affiliation{%
  \institution{Simulation \& Data Lab Neuroscience, J{\"u}lich Supercomputing Centre, Forschungszentrum J{\"u}lich}
  \country{Germany}
}
\email{k.singh@fz-juelich.de}

\author{Eric M{\"u}ller}
\orcid{0000-0001-5880-2012}
\affiliation{%
  \institution{Kirchhoff Institute for Physics, Ruprecht-Karls-Universit{\"a}t}
  \city{Heidelberg}
  \country{Germany}
}
\email{mueller@kip.uni-heidelberg.de}

\author{Eleni Mathioulaki}
\orcid{0009-0008-2019-7724}
\affiliation{%
  \institution{ATHENA Research \linebreak[2]{} \& Innovation Center}
  \city{Marousi}
  \country{Greece}}
\email{emathioulaki@athenarc.gr}

\author{Wouter Klijn}
\orcid{0000-0002-8996-488X}
\affiliation{%
  \institution{Simulation \& Data Lab Neuroscience, J{\"u}lich Supercomputing Centre, Forschungszentrum J{\"u}lich}
  \country{Germany}}
\email{w.klijn@fz-juelich.de}

\author{Lena Oden}
\orcid{0000-0002-9670-5296}
\affiliation{%
  \institution{Computer Engineering, FernUniversit\"{a}t in Hagen}
  \country{Germany}
}
\email{lena.oden@fernuni-hagen.de}

\renewcommand{\shortauthors}{Singh et al.}



\begin{abstract}
  Deploying complex, distributed scientific workflows across diverse \gls{hpc} sites is often hindered by site-specific dependencies and complex build environments.
  This paper investigates the design and performance of portable \gls{hpc} container images capable of encapsulating MPI- and CUDA-enabled software stacks without sacrificing bare-metal performance.
  This work is part of recent work performed within the EBRAINS Research Infrastructure, to evaluate the implementation of portable \gls{hpc} (Apptainer-based) container images targeting the \acrfull{esd} --- a Spack-based software ecosystem comprising approximately 80 top-level packages (and 800 dependencies).
  We evaluate a hybrid, PMIx-based containerization strategy using Apptainer that seamlessly bypasses the need for site-specific builds by dynamically leveraging host-level specialized hardware, such as network interfaces and GPUs, on two production HPC clusters: Karolina and Jureca-DC.
  We demonstrate the feasibility of building portable, MPI- and CUDA-enabled scientific software into container images that correctly leverage site-installed drivers and hardware to reproduce bare-metal communication behavior.
  Using communication microbenchmarks (e.g., OSU and NCCL) alongside performance metrics of applications from neuroscience, we measure and verify their performance against bare-metal deployments.
  Crucially, our verification approach extends beyond top-level runtime measurements;
  we highlight the analysis of underlying debug logs to actively detect misbehavior and misconfigurations, such as suboptimal transport pathways.
  Ultimately, this investigation demonstrates the feasibility of a simple and reproducible methodology for decoupling software environments from underlying infrastructures, paving the way for automated pipelines that ensure optimized, performance-verified execution across varied \gls{hpc} architectures.
\end{abstract}

\maketitle

\section{Introduction}

The deployment of complex, distributed scientific workflows across diverse \gls{hpc} platforms presents significant challenges regarding software portability and reproducibility.
Researchers frequently encounter conflicting version requirements, incompatible toolchains, and platform-specific build configurations.
While containerization technologies like Apptainer \cite{kurtzer2017singularity} have emerged as powerful tools to encapsulate software environments, reliably deploying heavily distributed, MPI- and CUDA-enabled applications within containers without sacrificing bare-metal performance remains a critical hurdle.
This paper presents our efforts to evaluate the implementation and performance of portable HPC containers for a representative subset of complex simulation tools from the \gls{esd} \cite{esd2026}.
The \gls{esd} is a community-driven, Spack-based ecosystem designed to unify analysis, modeling, and simulation software for brain research.
By building portable, MPI- and CUDA-enabled software directly into Apptainer images, we investigate whether these dependencies can be reliably containerized while correctly leveraging site-installed libraries. 
To rigorously evaluate this, we run communication microbenchmarks --- specifically OSU initialization, OSU latency, and NCCL tests --- to quantify container overhead, alongside neuroscience-specific scaling evaluations.
Beyond performance verification, a key objective of this work is the integration of these portable containers into existing production workflows.
For many EBRAINS users, a simple, easy-to-use environment is also a priority:
the \gls{esd} currently utilizes an automated, CI-driven build and test flow for service deployments, non-\gls{hpc} image builds and native site deployments.
The insights and verified methodologies from this investigation will be integrated directly into the \gls{esd}'s future portable \gls{hpc} container image build process. 
Ultimately, this will enable researchers to deploy an identical, performance-verified \gls{esd} environment not only across federated EBRAINS infrastructure but also on independent, external production \gls{hpc} systems.

\section{State of the Art}

The growing reliance on complex computational models has positioned software as a pillar of modern scientific research \cite{hong2025future}.
However, managing the lifecycle of scientific software presents persistent challenges regarding reproducibility, sustainability, and usability.
Scientific software is a multifaceted socio-technical system whose complexity, spanning engineering, governance, and infrastructure, is routinely overlooked in scientific practice \cite{hocquet2024software}.
A comprehensive survey of the technical barriers to reproducibility in scientific computing identifies the failure to encapsulate the entire computational environment as a root cause \cite{ivie2018reproducibility}:
software exists not merely as code but within a layered stack of operating systems, compilers, and library dependencies, any of which can silently alter or break results upon updates or platform changes \cite{hocquet2021epistemic}.
These environmental assumptions are rarely fully documented in publications \cite{howison2016software,schindler2022role}, and empirical studies confirm the scale of the problem:
attempts to reproduce published computational results succeed in only a fraction of cases \cite{stodden2018empirical}.
Even when code is available, only a minority of published computational papers can be successfully built and executed, with missing dependencies and build failures being among the most common barriers \cite{collberg2016repeatability,mangul2019challenges}.

Consequently, ensuring that research software can be reliably deployed across different systems without specialized domain knowledge remains a critical, ongoing challenge for the scientific community.

Migrating these workflows to \gls{hpc} environments introduces an additional layer of complexity.
\Gls{hpc} systems are characterized by specialized hardware architectures, high-speed interconnects, and restrictive, multi-tenant privilege models.
Deploying software on such infrastructure often leads to performance bugs if the tools are not perfectly tuned to the host environment \cite{thies2021rse,grannan2020understanding,azad2023empirical}.
To migrate the ``software chaos'' inherent in compiling tools against highly specific system libraries, package managers like Spack have become state of the art \cite{gamblin2015spack}.
Spack enables structured dependency management by allowing users to specify versions, compilers, and target microarchitectures, effectively automating the intricate build processes required for \gls{hpc}.

\paragraph{Containerization:}
Simultaneously, containerization has emerged as a primary strategy for achieving software portability.
Optimized container runtime engines, such as Apptainer, were explicitly designed to address the security model and architectural limitations of Docker in \gls{hpc} environments.
Containers allow researchers to encapsulate their software environment into a single image.
While recent surveys on containerization in \gls{hpc} demonstrate their efficacy in isolating software \cite{kurtzer2017singularity,zhou2022containerization,copik2025xaas}, achieving bare-metal performance --- particularly over high-speed interconnects and specialized accelerators --- remains an active area of investigation.

\paragraph{Dependency Management:}
Despite these advancements, comprehensive solutions that unify structured dependency management with automated containerization for entire software ecosystems remain scarce, with the \acrfull{e4s} being an notable exception \cite{willenbring2022impacting}.
While some recent reviews have examined broad software ecosystems \cite{moreau2023containers}, they generally lack a focus on combining rigorous package management (like Spack) with container deployment at scale.
Bridging this gap --- by uniting structured dependency management and software deployment with containerization to generate portable \gls{hpc} images --- represents a critical next step.

The investigation presented in this work serves as a foundational step toward integrating such capabilities into the \gls{esd}.
By evaluating the performance and feasibility of portable \gls{hpc} containers for a representative subset of tools prior to their full integration, we pave the way for a unified, performance-verified deployment pipeline for the entire \gls{esd} ecosystem.

\paragraph{Limits of Encapsulation}
While containers successfully isolate user-space software, they cannot encapsulate host-coupled infrastructure.
In HPC, this isolation boundary breaks down primarily at the accelerator and network levels.
For GPU workloads, a container image can bundle, e.g., the CUDA toolkit and runtime, but, as of today, it remains fundamentally dependent on the host kernel.
The container must dynamically bind-mount the host's proprietary device drivers (including user-space libraries such as \texttt{libcuda.so} that are tightly-coupled to kernel drivers), enforcing a strict \gls{abi} dependency across the container boundary.

Similarly, distributed execution topologies cannot be isolated.
To achieve MPI compatibility between containers and host systems, several alternative approaches rely on runtime translation layers or forwarding interfaces.
A containerized MPI stack must utilize the host's physical high-speed interconnects (e.g., InfiniBand) and synchronize with the host's resource manager (e.g., Slurm).
This necessitates an explicit protocol bridging the host and the container.
Typically, this is achieved via the Process Management Interface (PMIx), where the isolated MPI ranks act as clients querying the host-side Slurm server to resolve distributed endpoints.
Consequently, HPC containerization cannot rely on pure isolation;
it requires architectural strategies that explicitly manage these non-encapsulatable kernel and network interfaces.

Because of this tight coupling, maintaining MPI compatibility across the container boundary can be a challenge.
To address this, several state-of-the-art approaches employ runtime translation or forwarding layers.
Tools such as WI4MPI \parencite{leon2021fly} perform runtime translation between different MPI implementations (e.g., translating Open MPI calls to MPICH), while MPItrampoline \parencite{schnetter2022mpi} provides a forwarding MPI ABI that allows compiled applications to dynamically link to diverse host libraries at runtime.
While these solutions are highly valuable for deploying prebuilt binaries or legacy software, MPI 5 \cite{mpi5} provides a stable ABI definition and enables swapping out MPI libraries at runtime \cite{hammond2023mpi}.

\section{Limitations}

While this study demonstrates the feasibility and minimal overhead of containerizing complex neuroscience environments, several limitations exist within the current evaluation.

\paragraph{Scope of Evaluation and Hardware Constraints:}
First, the performance verification was conducted on a subset of the \gls{esd}, specifically focusing on communication microbenchmarks and isolated simulator scaling (Arbor \cite{abiakar2019arbor,hater_thorsten_ring_2025}, NEURON \cite{kumbhar2019coreneuron,hines1997neuron,neuron_ringtest_2026}) dependencies.
However, it does not yet capture the performance dynamics of the entire 80-package ecosystem.
We also observed a constant relative overhead for GPU-accelerated Arbor runs within Apptainer.
Because our current work focuses on runtime execution and scaling, a detailed investigation into the exact mechanisms, e.g., the influence of different userspace CUDA versions in bare-metal and containerized environments, is missing from this study and remains a key priority for future work.

\paragraph{Isolated Use Cases vs. Ecosystem Composability:}
Second, typical performance publications --- including the microbenchmarks in this study --- tend to focus on specific, isolated use cases.
However, the broader mission of the \gls{esd} is to enable new science by allowing researchers to seamlessly compose diverse tools into complex, multi-step workflows.
A limitation of our current benchmarking suite is that it does not yet evaluate the performance overhead of cross-tool interactions and coupled simulations within the containerized environment.
Future work will expand our \gls{ci} pipeline to include automated functional and performance testing of these composed workflows, facilitating the creation and modification of complex pipelines without degrading performance.

\paragraph{Automating Domain Expertise:}
Finally, deploying high-perfor\-mance containers typically requires deep \gls{hpc} domain knowledge to tune site-specific drivers and network transports.
While our containers successfully utilized site-installed libraries, detecting silent performance degradation --- such as a container falling back to a suboptimal network transport protocol --- currently requires expert manual review of trace logs.
To ensure true sustainability, reproducibility, and a ``useful'' deployment for non-expert users, we are actively expanding our deployment flow.
A key future output will be the integration of automated log parsing to proactively evaluate debug messages, immediately detecting and correcting suboptimal transport pathways without requiring user intervention.

\section{Solutions}

\subsection{Portable Containers for \gls{hpc}}

We adopt a \emph{PMIx-based hybrid} containerization strategy in which the container is entirely self-sufficient with respect to its MPI stack \parencite{castain2018pmix}.
The image carries a complete Open~MPI~5 installation compiled against its own internal PMIx, UCX, PRRTE, hwloc, and libevent libraries, with no dependency on any host-side MPI installation \cite{graham2006open,bernholdt2024taking,gabriel2004open}.
However, another option would be to make use of host libraries at runtime:
modern Open~MPI versions \cite{openmpi4} allow the underlying MPI library to be replaced at runtime, which is possible because Open~MPI has maintained a stable Application Binary Interface (ABI) since major version 4.
Furthermore, the new MPI-5 standard now explicitly provides a formal ABI definition \cite{hammond2023mpi}, ensuring robust, standardized interoperability across different implementations.

Containerized jobs are submitted to Slurm identically to native jobs, with the sole modification of specifying the PMIx wire-up protocol. 
The \texttt{-{}-mpi=pmix} flag instructs Slurm to act as the PMIx server, publishing process-map and endpoint information that the container's Open~MPI~5 runtime queries at start-up via the standard PMIx client interface. 

On the Karolina and Jureca-DC clusters used in this study, the container's UCX transport layer selects InfiniBand Verbs (\texttt{libib\-verbs}) for inter-node traffic and the POSIX shared-memory transport for intra-node communication, matching the fabric topology of both clusters without any site-specific tuning. For GPU-accelerated workloads, NCCL's topology detection handles GPU-to-GPU transfers, selecting NVLink where available and falling back to PCIe peer-to-peer access otherwise. The only site-specific requirement is that the host kernel exposes the InfiniBand device files (\texttt{/dev/infiniband/}) and that the container is launched with the \texttt{-{}-nv} flag (Apptainer), which exposes \texttt{/dev/nvidia*} devices and bind-mounts the basic NVIDIA libraries and CUDA libraries from the host.

\subsubsection{Portability and Reproducibility}

Because the entire MPI stack is encapsulated within the container image, the same image file can be transferred to any PMIx-capable Slurm cluster and executed without recompilation or site-specific tuning.
This satisfies a key reproducibility requirement for HPC benchmarking studies:
the software environment is immutable and version-pinned, and any performance difference between sites reflect hardware and network topology rather than software configuration
divergence.
PMIx support is present in modern Slurm \cite{schedmd2024mpi}, covering every major European HPC facility including Karolina, JUWELS~Booster, MareNostrum~5, LUMI, and Leonardo.

\subsection{Build Flow and Deployment}
To operationalize this at scale, we are integrating a specialized Slurm-based container build flow into our build and deployment tool for spack environments, \texttt{dedal}.

\subsubsection{Connectivity: Fetching and Building}
By separating the fetch and build stages for container construction, we can make use of \gls{hpc} resources for building software.
The \gls{ci} pipeline first fetches and builds base dependencies to construct a minimal base image for building.
Afterwards, Spack can identify and fetch all required sources.
Using the base image and the pre-fetched sources on the build\slash{}compute nodes, we can now build software in parallel, relying on Spack's parallel build support.

\subsubsection{Software Builds on HPC}
\label{sssec:sw_builds}
Because many compilation tools are highly sensitive to, e.g., file locking, latency and access time semantics of network-based file systems, the actual build is performed entirely within node-local file systems (typically in-memory using \texttt{/dev/shm}).
Once the Spack packages are built, the artifacts can be packed back into per-package images and stored on distributed file systems.
The complete Apptainer image can now be built by integrating the artifacts into the image (optionally dropping all build-time-only dependencies).
This workflow not only accelerates the \gls{ci} process but also fundamentally solves a typical \gls{hpc} infrastructure bottleneck:
by producing a limited number of files per package, it dramatically reduces inode exhaustion, which is typically a far stricter limitation on \gls{hpc} shared file systems than raw block storage capacity.

\section{HPC Systems and Environments}{\label{sec:systems}}

All benchmarks were conducted on two production HPC clusters: Karolina \cite{karolina}, operated by 
IT4Innovations (Czech Republic), and JURECA-DC \cite{jureca}, operated by Jülich Supercomputing 
Centre (Germany). Both systems were used to evaluate native execution against a containerized Apptainer image for all strong and weak scaling benchmarks.
Two container images were built, one for CPU-only workloads and one for GPU-accelerated workloads, with each image self-contained and transferred to both clusters without modification.

The container software toolchain versions listed in ~\cref{tab:software} reflect the environments available at the time the container images were built.
Since then, the native software stack on JURECA-DC has been updated and all host benchmarks were executed using the latest available toolchain on each cluster, while the container images remained unchanged.
This results in version differences in GCC, PMIx, and OpenMPI between the container and host environments.
The CPU container was additionally built on Rocky Linux~\texttt{10.1}, newer than  the host OS on either cluster.
That the same image executed successfully across both systems despite these discrepancies suggests that the container runtime provides sufficient isolation from the underlying host software stack.

For GPU-accelerated workloads, an additional consideration arose from CUDA version compatibility.
The GPU container was built against CUDA~\texttt{12.2}, which is lower than the versions available on either cluster (\texttt{12.4} and \texttt{13.0} respectively).
This was a deliberate decision to maximise portability across a broader range of HPC centres beyond those evaluated here, as a container built against an older CUDA version remains 
compatible with newer host drivers. Results on additional systems are not presented in this work.

\subsection*{Hardware}

Karolina compute nodes are each equipped with two AMD EPYC 7H12 processors (64 cores per socket, 128 cores per node). Inter-node communication uses InfiniBand HDR. For GPU workloads, each node hosts eight NVIDIA A100 GPUs interconnected via \gls{nv12}, providing a fully connected GPU topology within the node. Each pair of GPUs shares a dedicated Mellanox ConnectX HDR InfiniBand NIC, with NIC--GPU affinity determined by PCIe proximity (PXB --- it refers to a peer-to-peer connection topology where communication between GPUs traverses multiple PCIe internal switches) \label{sec:karolina-node}. 

JURECA-DC compute nodes are each equipped with two AMD EPYC 7742 processors (64 cores per socket, 128 cores per node), connected via InfiniBand HDR100 using NVIDIA Mellanox ConnectX-6 adapters.
For GPU workloads, each node carries four NVIDIA A100 GPUs connected via \gls{nv4}.
Unlike Karolina, where each GPU has a dedicated NIC at PXB proximity, the JURECA-DC node topology is heterogeneous:
only GPU1 and GPU2 have direct PIX-level affinity to NIC0 and NIC1 respectively. 
This asymmetry means that inter-node message injection latency and bandwidth are not uniform across all four GPUs, and collective operations that rely on all GPUs contributing equally to network traffic may be subject to contention on the NIC-proximate GPU paths \label{sec:jureca-node}.

\subsection*{Software Toolchain}{\label{tab:software}}

\cref{tab:software} summarizes the key software components used across the three 
environments, where \emph{RL} denotes Rocky Linux.
The container image versions were fixed at  build time and remained identical across both clusters, whereas the native environments reflect the module stacks available on each system.

\begin{table}[ht]
    \setlength{\tabcolsep}{3pt} 
	\centering
	\caption{Software toolchain comparison across native and containerized environments. RL~=~Rocky Linux.}
	\label{tab:software}
	\small
	\begin{tabular}{lcccc}
		\toprule
		\textbf{Component} & \textbf{Karolina} & \textbf{JURECA-DC} & \textbf{Container} & \textbf{Container}   \\
                           &                   &                    & \textbf{(CPU)}     & \textbf{(GPU)}       \\
		\midrule
		OS        & RL \texttt{8.10}       & RL \texttt{9.7}        & RL \texttt{10.1}   & RL \texttt{9.7}    \\
		GCC       &    \texttt{14.3.0}     &    \texttt{14.3.0}     &    \texttt{14.3.1} &    \texttt{12.2.1} \\
		CUDA      &    \texttt{12.4.0}     &    \texttt{13.0}       & ---                &    \texttt{12.2}   \\
		UCX       &    \texttt{1.19.0}     &    \texttt{1.19.0}     &    \texttt{1.19.0} &    \texttt{1.19.0} \\
		PMIx      &    \texttt{5.0.8}      &    \texttt{5.0.8}      &    \texttt{5.0.3}  &    \texttt{5.0.3}  \\
		OpenMPI   &    \texttt{5.0.8}      &    \texttt{5.0.8}      &    \texttt{5.0.5}  &    \texttt{5.0.5}  \\
		Apptainer &    \texttt{1.4.2}(el8) &    \texttt{1.4.5}(el9) & ---                & ---                \\
		\bottomrule
	\end{tabular}
\end{table}

\section{Benchmarking}
To compare the performance of native and containerized execution, we conducted microbenchmarks to capture the overheads introduced by the container. Startup performance was measured using the OSU Micro-Benchmarks initialization test (\texttt{OSU\_init}) \cite{osu_benchmarks,panda_mvapich_2021}. Network latency and communication costs were evaluated via \texttt{OSU\_latency}, allowing us to assess whether containerization introduces measurable penalties compared to native high-performance interconnect access. GPU communication efficiency was examined using \texttt{NCCL} tests, focusing on bandwidth and latency during inter- and intra-node data transfers to determine whether containers affect direct GPU access or PCIe/NVLink utilization \cite{nccl_tests} \cite{nvidia_nccl}.

In addition to these microbenchmarks, we conducted neuroscience-specific benchmarks using the \texttt{Arbor} simulator (CPU and GPU) \parencite{abiakar2019arbor} and the \texttt{NEURON} simulator (CPU) to represent real-world workload performance \parencite{hines1997neuron}. These benchmarks capture the behavior of computational neuroscience applications, providing insight into how containerization affects realistic simulation workloads.

\subsection{Micro-Benchmarks}

\subsubsection{\texttt{OSU Init}} The \texttt{osu\_init} benchmark quantifies the wall-clock time required to execute \texttt{MPI\_Init()} across all participating ranks. This metric captures the cumulative latency of transitioning the MPI runtime into a communicable state. 
The initialization phase encompasses critical operations, including:
\begin{itemize}
    \item PMI/PMIx key-value exchanges for rank synchronization.
    \item Discovery of transport mechanisms and underlying network fabrics.
    \item Allocation of shared memory segments and distribution of endpoint addresses.
    \item Configuration of internal runtime data structures.
\end{itemize}

While native installations benefit from direct integration with the host OS and system libraries, execution within an Apptainer container introduces potential overhead.
This overhead primarily stems from operations needed for the isolation of namespaces.
When scaling to high process counts, the per-process or per-node replication of the setup routines can result in observable performance degradation during the initialization window compared to native execution.

\begin{figure}[htb]
    \centering
    \includegraphics[width=\columnwidth]{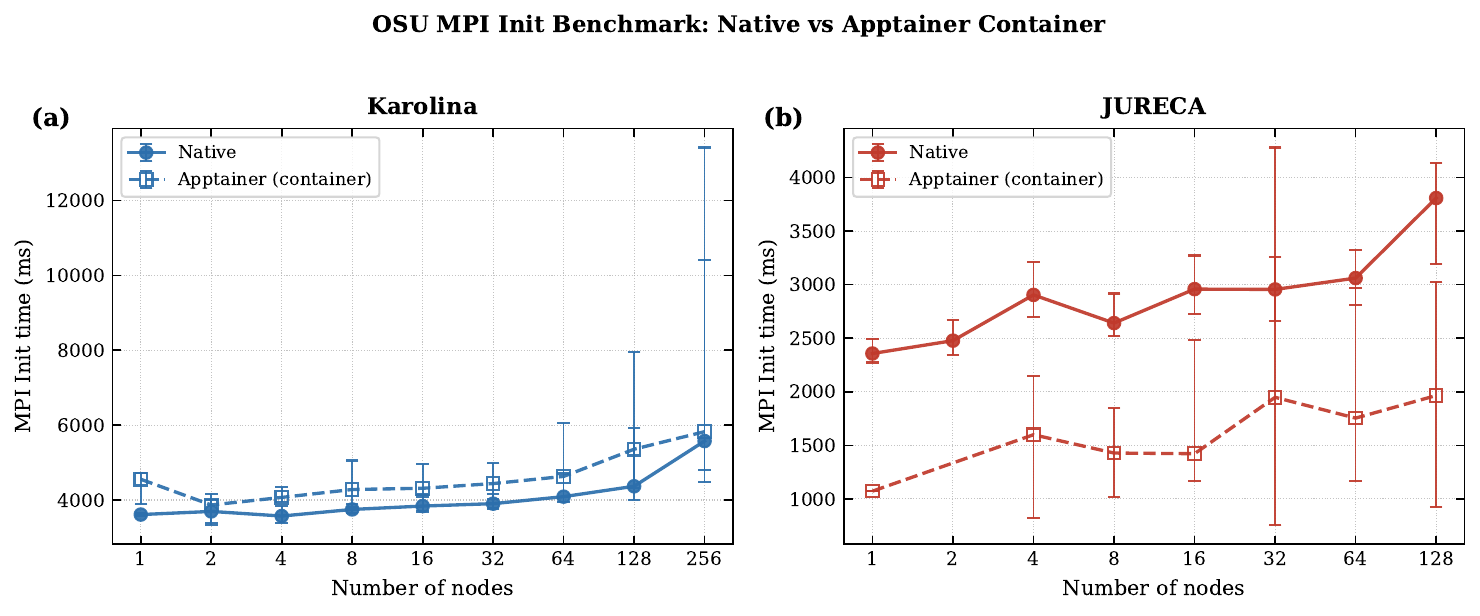}
    \caption{\normalfont \texttt{osu\_init} benchmark results comparing MPI initialization time for native and Apptainer container execution on Karolina (a) and JURECA (b). Error bars represent the minimum and maximum observed initialization times across all runs. On Karolina, the container incurs a consistently higher initialization overhead than native execution, with the gap widening at 256 nodes. On JURECA, the container exhibits significantly lower initialization times across all node counts, suggesting a leaner MPI bootstrap path within the containerized environment. Lower is better.
    }
    \label{fig:osuinit}
\end{figure}

The \texttt{osu\_init} benchmarks reveal highly system-dependent behavior regarding container initialization overhead. On Karolina (\cref{fig:osuinit}a), containerized execution consistently lags behind host-native performance, with both the latency gap and performance variability increasing as the node count scales. This suggests that the additional namespace isolation layers in Apptainer may exacerbate PMIx bootstrap latencies at scale on this specific architecture.

In contrast, the Jureca system (\cref{fig:osuinit}b) exhibits a significant performance advantage for containerized workloads, with initialization times reduced by approximately \SI{50}{\percent} across all tested scales. This indicates that the containerized environment likely benefits from a more streamlined MPI runtime configuration, effectively bypassing costly host-level device discovery and transport probing routines.

The contrasting behavior between these two systems highlights that the container overhead is not a universal constant;
instead, initialization performance is a function of how the containerized runtime interacts with the specific transport stack and process management interface of the host system.

\subsubsection{\texttt{OSU latency}} We evaluate \texttt{osu\_latency} across two distinct configurations to characterize different aspects of the communication overhead:

\begin{itemize}
    \item \textbf{Intra-node:} Both MPI processes are mapped to the same physical node. In this configuration, communication typically leverages shared-memory transports. The resulting latency is primarily a function of the shared-memory subsystem efficiency and the operating system's scheduling behavior.
    
    \item \textbf{Inter-node:} The MPI processes are distributed across separate physical nodes, forcing communication to traverse the cluster interconnect. In this setting, latency measurements reflect the performance of the MPI transport layer in conjunction with the underlying network fabric.
\end{itemize}

\begin{figure}[htb]
	\centering
	\includegraphics[width=\columnwidth]{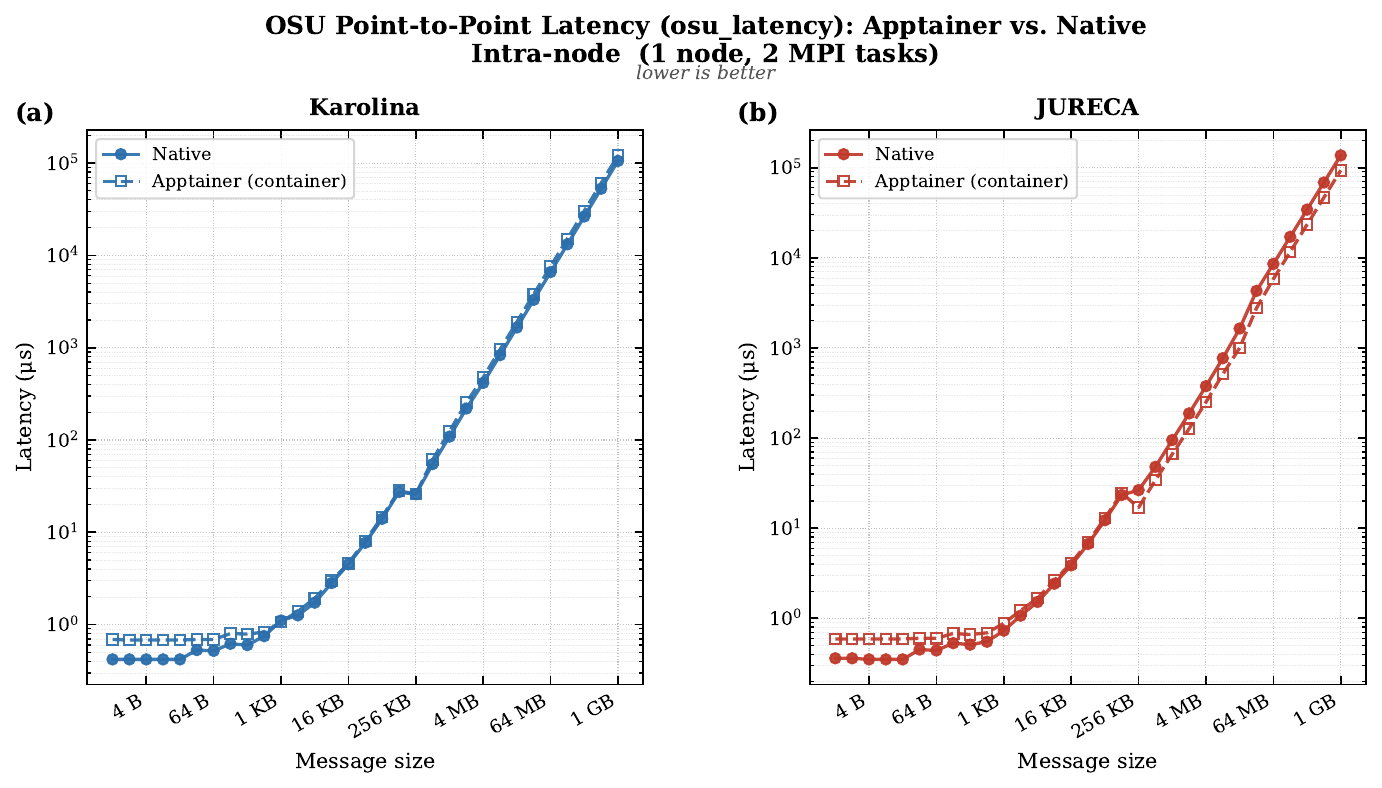}
	\caption{\normalfont OSU point-to-point latency (osu\_latency) for intra-node communication (1 node, 2 MPI tasks on the same node) on (a)~Karolina and (b)~JURECA. Latency is plotted as a function of message size for Apptainer (container) and native bare-metal execution. Both axes are logarithmic. Lower is better.}
	\label{fig:osulatencyfigure}
\end{figure}

\begin{figure}[htb]
	\centering
	\includegraphics[width=\columnwidth]{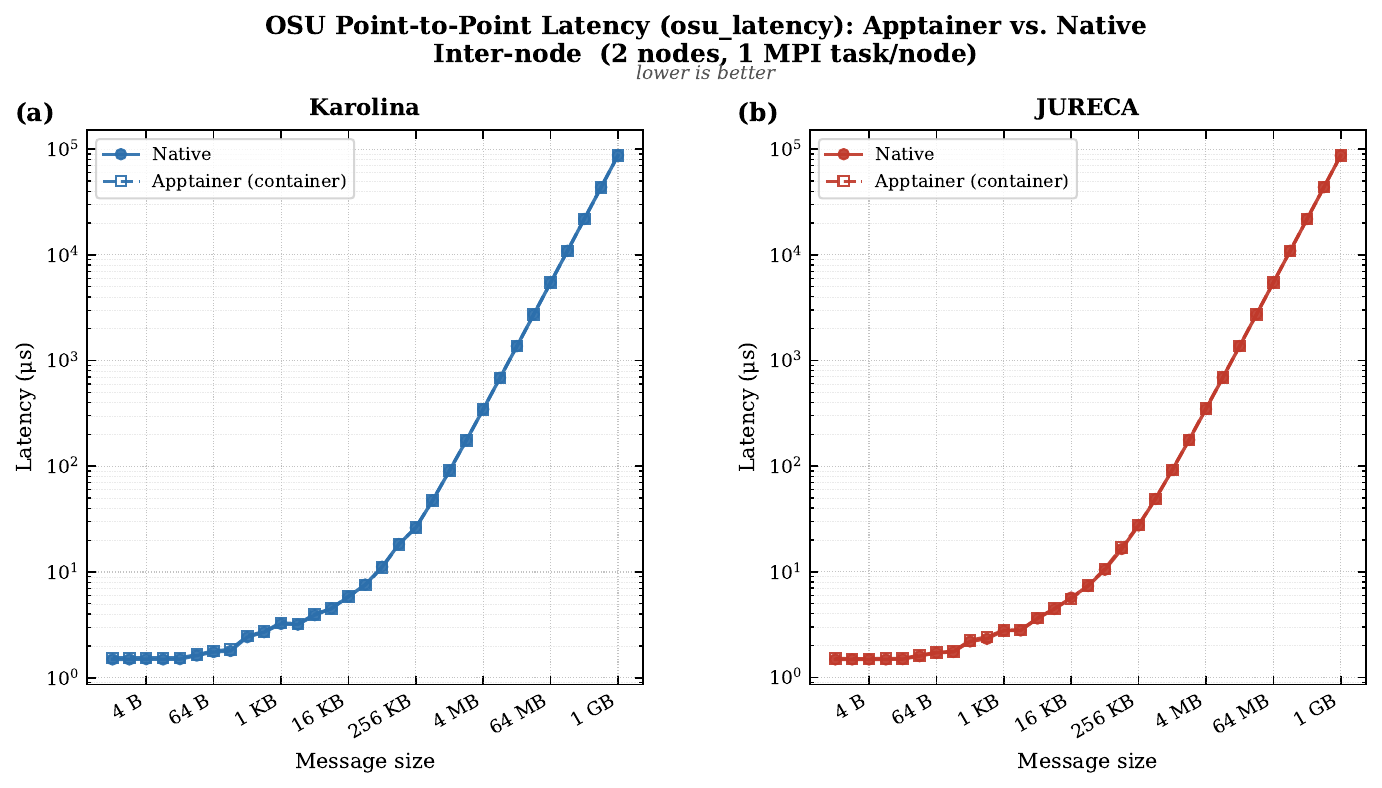}
	\caption{\normalfont OSU point-to-point latency (osu\_latency) for inter-node communication (2 nodes, 1 MPI task per node) on (a)~Karolina and (b)~JURECA. Latency is plotted as a function of message size for Apptainer (container) and native bare-metal execution. Both axes are logarithmic. Lower is better.}
	\label{fig:osulatencyInter}
\end{figure}

The benchmarking results (\cref{fig:osulatencyfigure} and \cref{fig:osulatencyInter}) across the Karolina and JURECA clusters confirm that Apptainer introduces negligible overhead across all message size regimes.
For small~messages ($\le \SI{1}{\kibi\byte}$), the absolute overhead is strictly sub-microsecond; intra-node communication shows a mean latency increase of approximately \SI{0.19}{\micro\second}, while inter-node overhead remains even lower at less than \SI{0.05}{\micro\second}. 
In the medium message range (\SIrange{1}{128}{\kibi\byte}), the performance delta remains minimal, with absolute overheads typically staying below \SI{0.5}{\micro\second}, ensuring that latency-sensitive synchronization steps in neuroscience simulations are not bottlenecked by the container layer. 
Finally, for large messages ($> \SI{128}{\kibi\byte}$), the communication is dominated by bandwidth rather than software overhead. In these cases, Apptainer performs nearly identically to the native environment for inter-node transfers, while intra-node results show that any performance variation is well within the expected margin of measurement noise or system jitter. 
Consequently, for the majority of neuroscience-related workloads, which rely on either inter-node scaling or large-scale data transfers, the latency penalty attributable to containerization is negligible.

\subsubsection{Performance Evaluation via NCCL-tests}

The \texttt{nccl-tests} suite, specifically the \texttt{all\_reduce\_perf} benchmark, serves as a
standard for validating the efficiency of the NVIDIA Collective Communications Library (NCCL) by
measuring the bus bandwidth of the \textit{AllReduce} collective operation. We evaluate two
distinct configurations to characterise different aspects of the communication stack:

\begin{itemize}
    \item \textbf{Single-node:} All GPU ranks reside on the same physical node, and communication
    exclusively uses the intra-node NVLink fabric. The objective is to validate whether the
    containerised environment introduces any overhead, such as misconfigured shared memory
    (\texttt{/dev/shm}) or restricted GPU device access.

    \item \textbf{Two-node:} GPU ranks are distributed across two physical nodes, forcing all
    collective traffic to traverse the InfiniBand fabric via GPUDirect RDMA\@. This setting
    directly exposes the inter-node injection bandwidth and quantifies any penalty introduced by
    the container on the network transport path.
\end{itemize}

\begin{figure}[htb]
    \centering
    \includegraphics[width=.9\columnwidth]{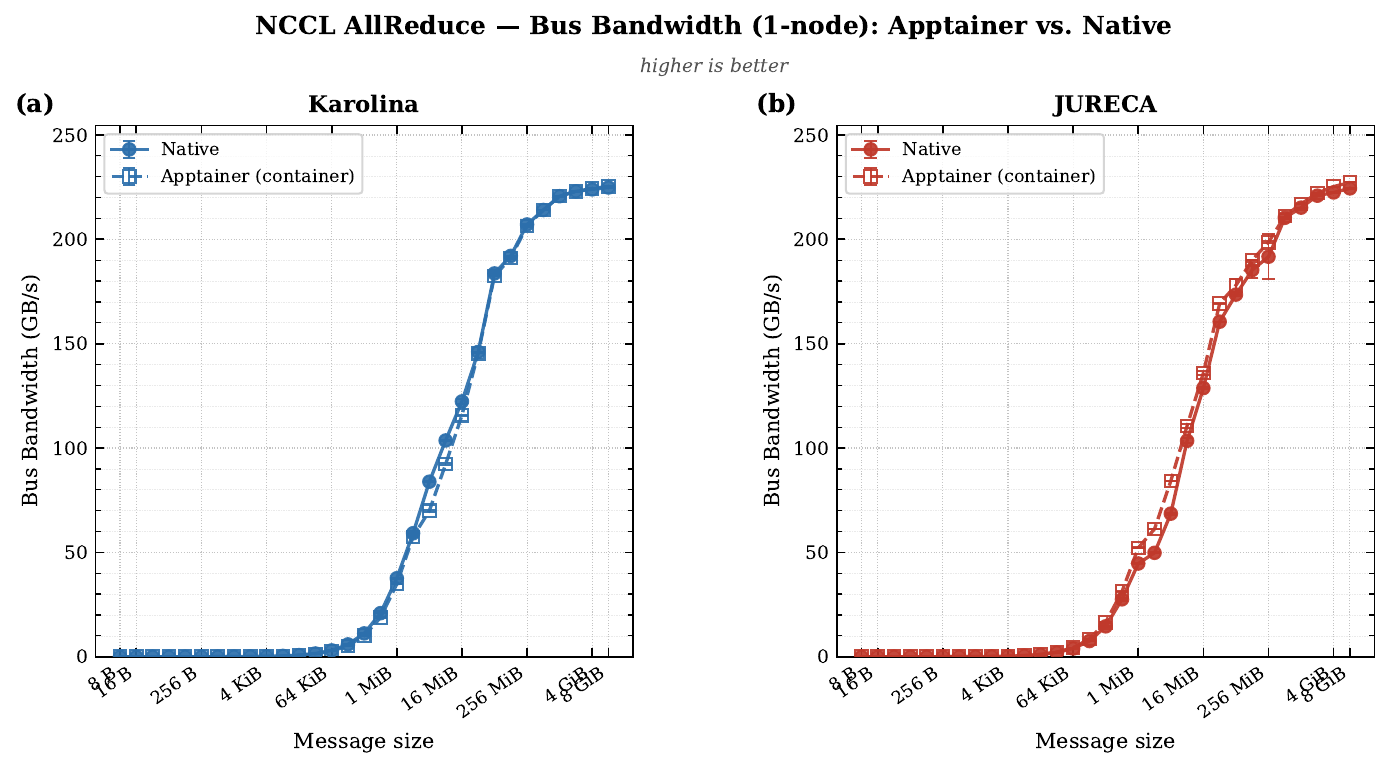}
    \caption{\normalfont NCCL \textit{AllReduce} bus bandwidth as a function of message size for a
        single-node configuration, comparing native execution and Apptainer containerisation.
        \textbf{(a)}~Karolina: 8~GPUs interconnected via NV12 NVLink bonds, peak bus bandwidth
        $\approx$\,\SI{225}{\giga\byte\per\second}.
        \textbf{(b)}~JURECA: 4~GPUs interconnected via NV4 NVLink bonds, peak bus bandwidth
        $\approx$\,\SI{225}{\giga\byte\per\second}.
        On both systems, native and Apptainer results are indistinguishable across all message
        sizes, with peak bus bandwidth deviating by at most \SI{1.3}{\percent}.
        Markers show the mean of two benchmark runs; error bars indicate the half-range between
        runs. Higher is better.}
    \label{fig:ncclcomparisonresults1node}
\end{figure}

\begin{figure}[htb]
    \centering
    \includegraphics[width=0.9\columnwidth]{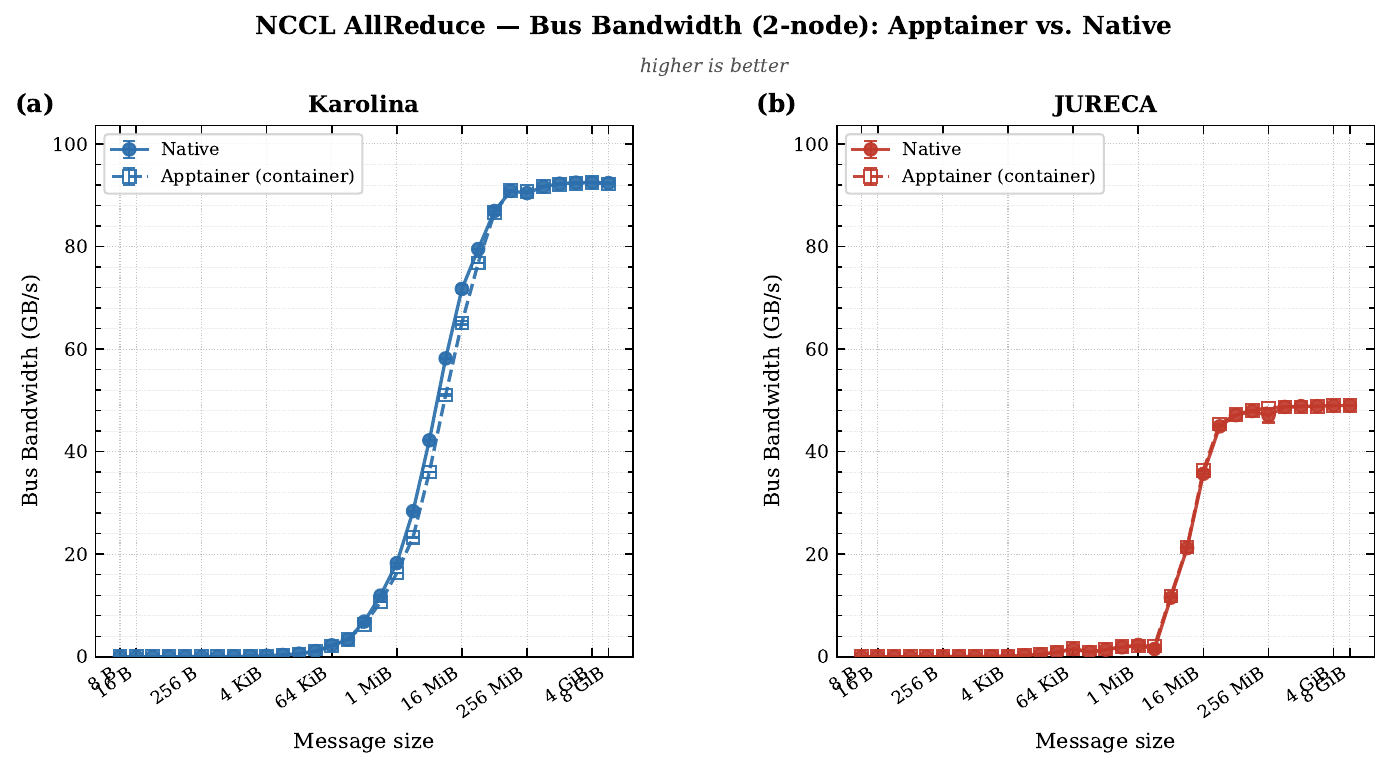}
    \caption{\normalfont NCCL \textit{AllReduce} bus bandwidth as a function of message size for a
        two-node configuration, comparing native execution and Apptainer containerisation.
        \textbf{(a)}~Karolina: peak inter-node bus bandwidth \SI{92.5}{\giga\byte\per\second},
        sustained by 4~InfiniBand NICs per node each connected to a dedicated GPU pair.
        \textbf{(b)}~JURECA: peak inter-node bus bandwidth \SI{49.0}{\giga\byte\per\second},
        limited by only 2~NICs per node.
        The $\approx$\,\SI{2}{\times} bandwidth difference between the systems reflects their
        NIC-to-GPU topology rather than any container effect: native and Apptainer results agree
        to within \SI{0.09}{\percent} on Karolina and \SI{0.01}{\percent} on JURECA across all
        message sizes. Higher is better.}
    \label{fig:ncclcomparisonresults2nodes}
\end{figure}

The benchmarking results (\cref{fig:ncclcomparisonresults1node} and
\cref{fig:ncclcomparisonresults2nodes}) confirm that Apptainer introduces negligible overhead
across all message size regimes and both node configurations.

For the \textbf{single-node} case, both systems saturate their respective NVLink fabrics at
approximately \SI{225}{\giga\byte\per\second} for large messages ($\geq$\,\SI{4}{\gibi\byte}),
with native and Apptainer results agreeing to within \SI{0.24}{\percent} on Karolina and
\SI{1.29}{\percent} on JURECA — well within run-to-run variability. At the \SI{8}{\byte}
baseline, small-message latency is \SI{36.6}{\micro\second} (native) versus
\SI{41.1}{\micro\second} (Apptainer) on Karolina, and \SI{24.1}{\micro\second} versus
\SI{21.4}{\micro\second} on JURECA; the slight container advantage on JURECA confirms these
differences reflect noise rather than systematic overhead.

For the \textbf{two-node} case, the transition to inter-node communication reduces peak bus
bandwidth substantially — from $\approx$\,\SI{225}{\giga\byte\per\second} to
\SI{92.5}{\giga\byte\per\second} on Karolina (a \SI{59}{\percent} reduction) and to
\SI{49.0}{\giga\byte\per\second} on JURECA (a \SI{78}{\percent} reduction). The
$\approx$\,\SI{2}{\times} difference between the two systems reflects their NIC-to-GPU topology,
as discussed in \cref{sec:systems}. Critically, this bandwidth gap is a hardware topology effect
and not a container artefact: native and Apptainer bus bandwidths agree to within
\SI{0.09}{\percent} on Karolina and \SI{0.01}{\percent} on JURECA, confirming that GPUDirect
RDMA operates correctly inside the Apptainer container on both systems.

Consequently, for GPU-accelerated neuroscience workloads that rely on NCCL collectives --- whether
for intra-node gradient aggregation or inter-node parameter synchronisation --- Apptainer
containerisation imposes no measurable performance penalty across either the bandwidth-bound
large-message or the latency-bound small-message regime.

\subsection{Application Benchmarks: Neuroscience Workloads}
To complement the MPI micro-benchmarks, we evaluated container performance using the Arbor \cite{abiakar2019arbor} and NEURON \cite{hines1997neuron} neuroscience simulators, targeting both scaling behaviour and cross-site portability. The container image was built once and executed without modification on both Karolina and JURECA, validating that the PMIx-based container approach transfers across clusters without recompilation or site-specific tuning. NEURON results are limited to CPU execution; a bare-metal comparison for GPU-accelerated NEURON was not feasible within the available compute allocation and is deferred to future work.
	
\subsubsection{Arbor Ring Benchmark (CPU)}
The arbor ring network benchmark constructs a synthetic spiking neural network of $N$ morphologically detailed cable cells arranged in a unidirectional ring topology, where each cell $i$ receives a single excitatory synaptic connection from its predecessor $i-1$ (modulo N) \cite{hater_thorsten_ring_2025} . Each cell is modelled using a Hodgkin-Huxley soma with passive dendritic compartments; an external stimulus fires cell 0, and the resulting action potential propagates sequentially around the ring via synaptic transmission with a fixed axonal delay.

The simulation is implemented using Arbor, which distributes cells evenly across MPI ranks and advances the network through a bulk-synchronous scheme: at each time step, all ranks integrate their local cell dynamics independently before exchanging generated spikes via a global \texttt{MPI\_Allgather} collective.

\begin{figure}[htb]
	\centering
	\includegraphics[width=\columnwidth]{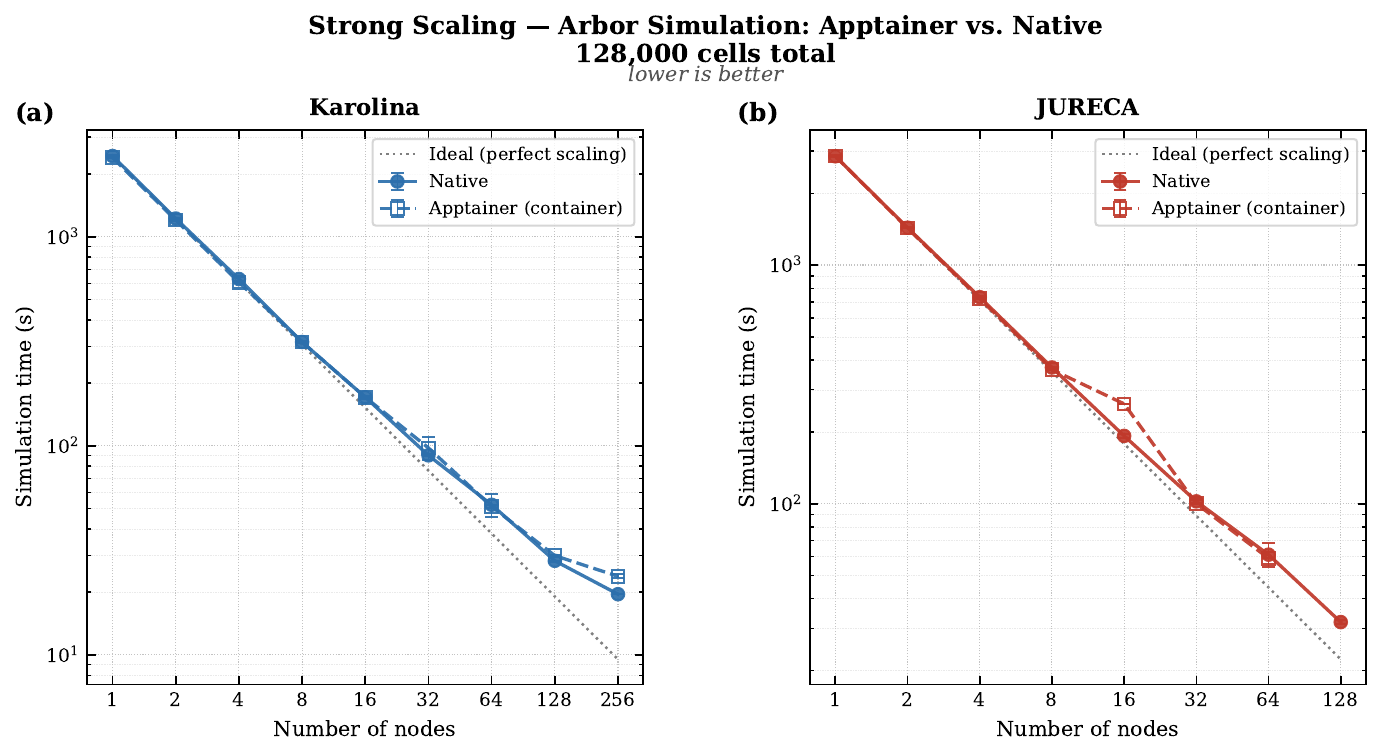}
	\caption{ \normalfont Strong scaling performance of the Arbor ring network benchmark (\num{128000} cells on (a) Karolina and (b) JURECA. Simulation time is plotted against the number of nodes for Apptainer (container) and native bare-metal execution. The dotted line represents ideal perfect scaling. Error bars denote the standard deviation across repeated runs. Lower is better.}
	\label{fig:ArborCpuSS}
\end{figure}

\paragraph{Strong Scaling (CPU)}
The strong scaling results for the Arbor ring network benchmark
($N = \num{128000}$ cells, fixed problem size) demonstrate that Apptainer
and native bare-metal execution achieve broadly comparable parallel
efficiency across both clusters, as shown in \cref{fig:ArborCpuSS}.

On Karolina (\cref{fig:ArborCpuSS}a), native execution reduces
simulation time from \SI{2435}{\second} at 1 node to
\SI{28.2}{\second} at 128 nodes (\SI{67.5}{\percent} efficiency), while the container achieves \SI{2391}{\second} and \SI{29.8}{\second}
respectively (\SI{62.6}{\percent} efficiency);
the two environments remain within $\sim\SI{5}{\percent}$ of each other across the entire 1--64 node range.

On JURECA (\cref{fig:ArborCpuSS}b), both environments exhibit consistent scaling behaviour across the 1--64 node range, with the container reaching \SI{98.0}{\percent} efficiency against \SI{95.9}{\percent} for native;
an outlier at 16 nodes ($+\SI{36}{\percent}$ overhead) is not reproduced at neighbouring node counts and is not considered a systematic effect.

\begin{figure}[htb]
	\centering
	\includegraphics[width=\columnwidth]{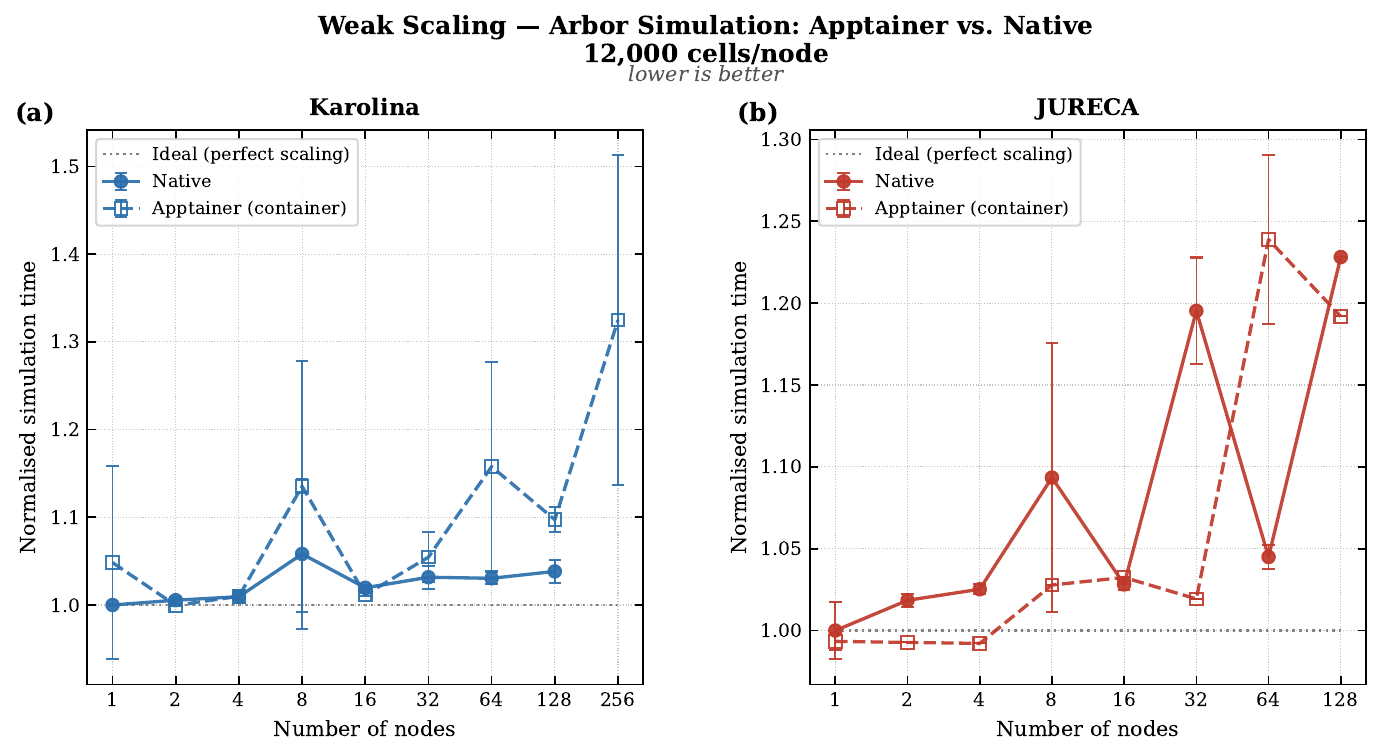}
	\caption{ \normalfont Weak scaling performance of the Arbor ring network benchmark (\num{12000} cells-per-node on (a) Karolina and (b) JURECA.  Simulation time is normalised to the single-node native baseline; a value of \num{1.0} (dotted line) indicates ideal scaling with no overhead as the problem size and node count grow proportionally. Error bars denote the standard deviation across repeated runs. Lower is better.
	}
	\label{fig:ArborCpuWS}
\end{figure}

\paragraph{Weak Scaling (CPU)} 
The weak scaling results (\num{12000} cells per node, growing problem size) are presented in \cref{fig:ArborCpuWS}. On Karolina (\cref{fig:ArborCpuWS}a), the native installation maintains near-ideal scaling throughout, with simulation time rising by no more than \SI{5.8}{\percent} above the single-node baseline across the entire 1--128 node range; the container follows the same trend closely, staying within \SI{5.5}{\percent} of baseline up to 32 nodes, though simulation times at 8 nodes ($+\SI{13.5}{\percent}$) and 64 nodes ($+\SI{15.8}{\percent}$) indicate occasional variance, likely attributable to system jitter given the non-monotonic pattern.

On JURECA (\cref{fig:ArborCpuWS}b), both environments exhibit
comparable scaling up to 16 nodes, remaining within \SI{3.2}{\percent} of
baseline;
beyond this point both curves show elevated variance, with
the native run reaching $+\SI{19.5}{\percent}$ at 32 nodes before recovering to $+\SI{4.5}{\percent}$ at 64 nodes, and the container reaching $+\SI{29.0}{\percent}$ at 64 nodes --- a pattern present in both environments that is consistent with cluster-level scheduling noise rather than a systematic containerization overhead.

Overall, these results confirm that Apptainer introduces no systematic
degradation in parallel efficiency, and that the dominant source of
efficiency loss at high node counts --- the growing fraction of wall-clock
time consumed by \texttt{MPI\_Allgather} spike-exchange collectives as
the per-node workload shrinks — affects both environments equally.

\subsubsection{Neuron CPU} We evaluate the parallel scaling behaviour of NEURON using the ringtest benchmark. The benchmark constructs a network of 256 independent ring circuits, each consisting of Hodgkin--Huxley cells connected in a unidirectional chain (similar to the arbor ring network) \cite{neuron_ringtest_2026}.

All runs used NEURON v9.0.1 with a biological simulation time of \SI{100}{\milli\second} and a fixed time step of \SI{0.025}{\milli\second} (\num{4000} integration steps), without CoreNEURON, relying solely on NEURON's native CPU solver. 

\begin{figure}[htb]
	\centering
	\includegraphics[width=\columnwidth]{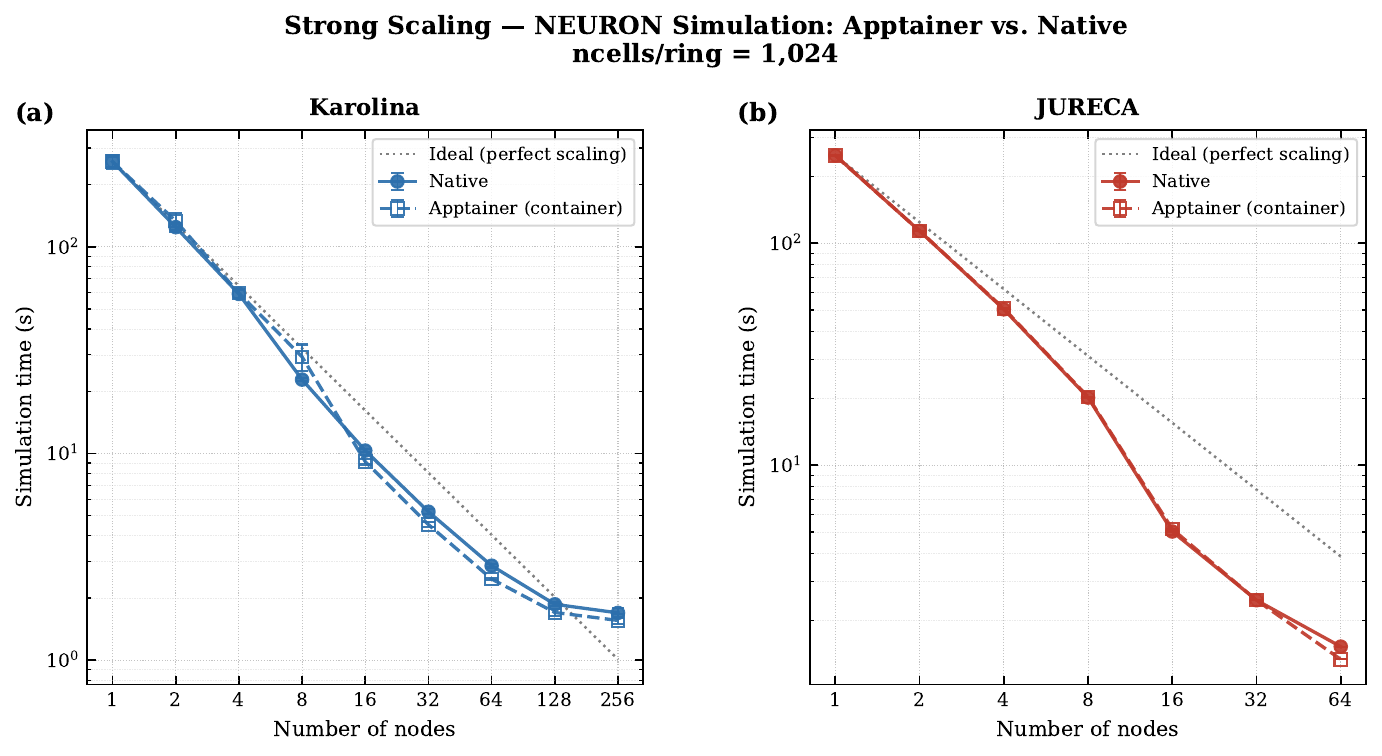}
	\caption{ \normalfont Strong scaling of the NEURON ring network benchmark (CPU) on
    (a)~Karolina and (b)~JURECA-DC. The total problem size is fixed at
    1,024~cells distributed across 256~rings (4~cells/ring), and the number
    of nodes is doubled from 1 to 8. Simulation wall-clock time is plotted
    on a log--log scale for Apptainer (container) and native bare-metal
    execution.
    The dashed black line denotes ideal strong scaling anchored to the
    1-node native mean. Lower is better.
    }
	\label{fig:neuron_strong}
\end{figure}

\paragraph{Strong Scaling (CPU)}
Strong-scaling performance was assessed by fixing the network at \num{1024} cells per ring across all 256 rings and increasing the node count from 1 to 256 nodes on Karolina and from 1 to 64 nodes on JURECA (\cref{fig:neuron_strong}).
On Karolina, the wall-clock time falls from approximately \SI{250}{\second} on a single node to around \SI{1.5}{\second} at 256 nodes. On JURECA, the runtime decreases from approximately \SI{248}{\second} on one node to between \SIrange{1.3}{1.5}{\second} at 64 nodes. Both systems achieve super-linear speedup at higher node counts.
Across the full node range and on both platforms, the native and Apptainer curves are nearly indistinguishable, with any difference falling within the run-to-run variability, confirming that containerisation imposes no measurable overhead on strong-scaling performance.

\begin{figure}[htb]
	\centering
	\includegraphics[width=\columnwidth]{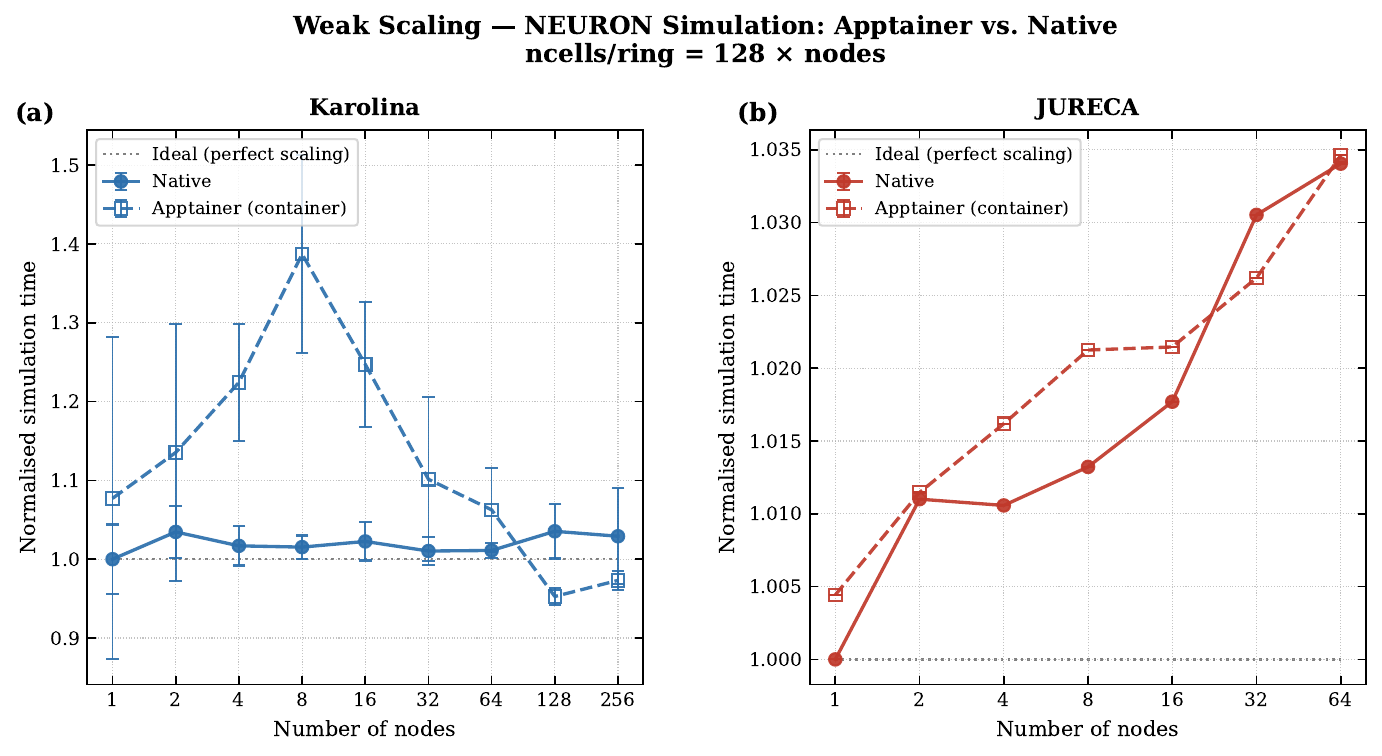}
	\caption{\normalfont Weak scaling of the NEURON ring network benchmark (CPU) on
    (a)~Karolina and (b)~JURECA-DC. The problem size grows proportionally
    with the number of nodes, so that the per-node workload remains
    constant. Simulation time is normalised to the 1-node.
    Results are shown for Apptainer (container) and native bare-metal
    execution; error bars show the standard deviation over repeated runs.
    Lower is better.
    }
	\label{fig:neuron_weak}
\end{figure}

\paragraph{Weak Scaling (CPU)}
For the weak-scaling study, the number of cells per ring was scaled proportionally with the node count as $\mathrm{ncells/ring} = 128 \times N$,
where $N$ is the number of nodes, while the number of rings was held constant at 256.
Runtimes are normalised to the single-node native execution and shown in \cref{fig:neuron_weak}.

On Karolina, the native environment maintains a normalised runtime close to \num{1.0} across all node counts up to 256 nodes, demonstrating excellent weak-scaling efficiency;
the Apptainer runs exhibit larger variance, particularly at 8 nodes, but converge towards the native performance at higher node counts.
On JURECA, both environments rise only gradually from 1.0 at one node to approximately \num{1.035} at 64 nodes, equivalent to a weak-scaling efficiency above \SI{96}{\percent} throughout the entire range.
The close agreement between native and container runtimes on both systems confirms that the Apptainer container layer introduces no significant communication or synchronisation overhead as the problem size scales with the node count.

\subsubsection{Arbor GPU} 

The GPU benchmarks were also conducted on Karolina and JURECA HPC clusters. Karolina is equipped with 8~GPUs per node whereas  JURECA provides 4~GPUs per node.

\begin{figure}[htb]
	\centering
	\includegraphics[width=\columnwidth]{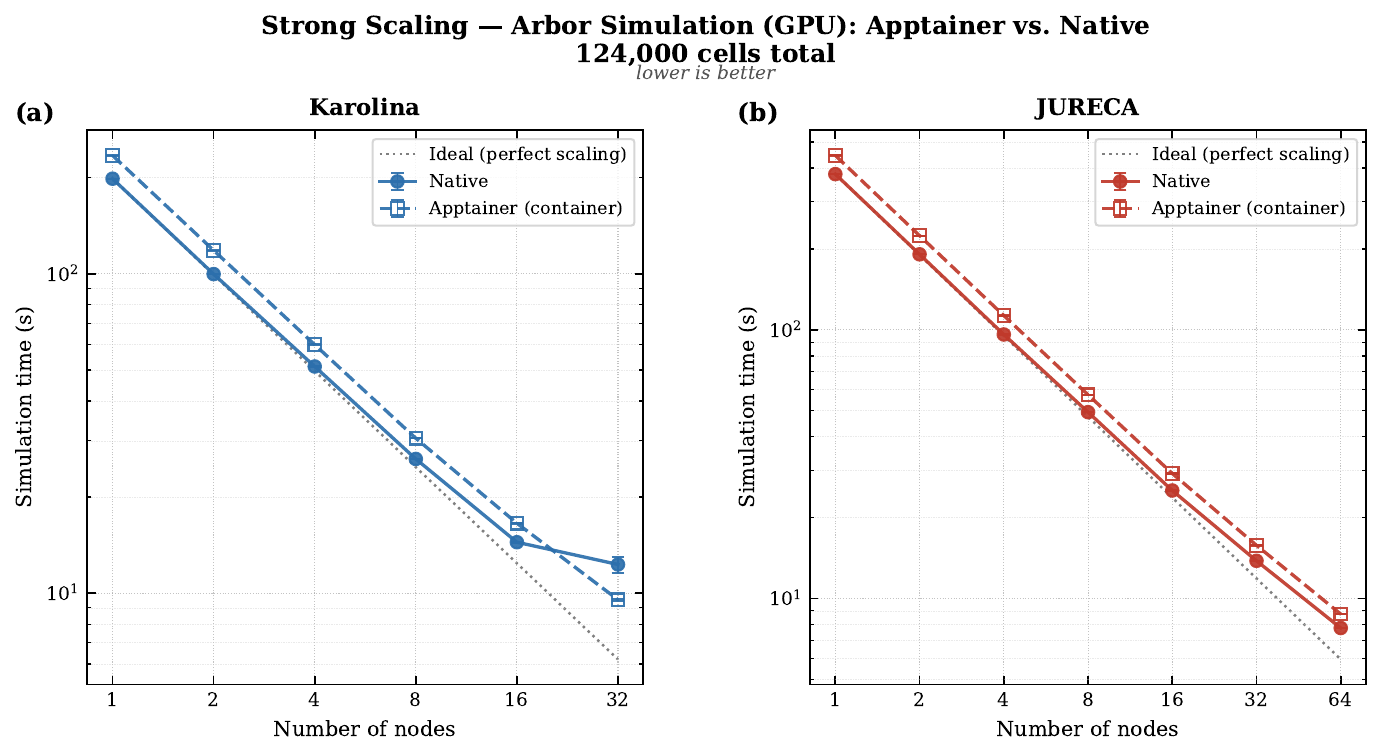}
	\caption{\normalfont Strong scaling performance of the Arbor ring network benchmark
		(GPU, $N = \num{124000}$ cells, 10 synapses per cell,
		$\delta_{\min} = \SI{5}{\milli\second}$, $T = \SI{200}{\milli\second}$) on
		(a)~Karolina and (b)~JURECA.
		Simulation time is plotted against the number of nodes for Apptainer
		(container) and native bare-metal execution.
		The dotted line indicates ideal perfect scaling.
		Error bars denote the standard deviation across repeated runs.
		Lower is better.}
	\label{fig:ArborGpuSS}
\end{figure}

\paragraph{Strong Scaling (GPU)} The GPU strong scaling results
($N = \num{124000}$ cells, fixed problem size) are presented in
\cref{fig:ArborGpuSS}. Unlike the CPU results, where no systematic container overhead was observed, the GPU results reveal a consistent overhead of Apptainer relative to native execution across both clusters. On Karolina (\cref{fig:ArborGpuSS}a), the container is
\SI{18.2}{\percent} slower than native at 1 node (\SI{234.8}{\second} vs.\ \SI{198.6}{\second}), and this overhead persists across the $1--16$ node range (\SIrange{14.7}{18.7}{\second}), indicating a fixed per-node cost attributable to the container layer rather than a communication-induced penalty that would grow with node count.

The parallel efficiency of both environments remains comparable up to 16 nodes, reaching \SI{85.9}{\percent} (native) and \SI{88.6}{\percent} (container);
the anomaly at 32 nodes --- where the container records a lower simulation time than native (\SI{9.5}{\second} vs.\ \SI{12.3}{\second})— is not reproducible across the scaling curve and is attributed to
system noise on that particular run rather than a systematic effect.

On JURECA (\cref{fig:ArborGpuSS}b), the pattern is also consistent: the container overhead decreases monotonically from \SI{17.6}{\percent} at 4 nodes to \SI{12.7}{\percent} at 64 nodes as the
communication fraction grows and the startup cost becomes relatively smaller, while parallel efficiency tracks closely between the two environments at every measured point, reaching \SI{76.4}{\percent} (native) and \SI{79.6}{\percent} (container) at 64 nodes.

Overall, the strong scaling results indicate that Apptainer introduces a constant relative
overhead of \SIrange{12}{19}{\percent} on GPU hardware for Arbor, 
though its precise origin requires further investigation.
In particular, we did not evaluate the influence of CUDA versions.

\begin{figure}[htb]
	\centering
	\includegraphics[width=\columnwidth]{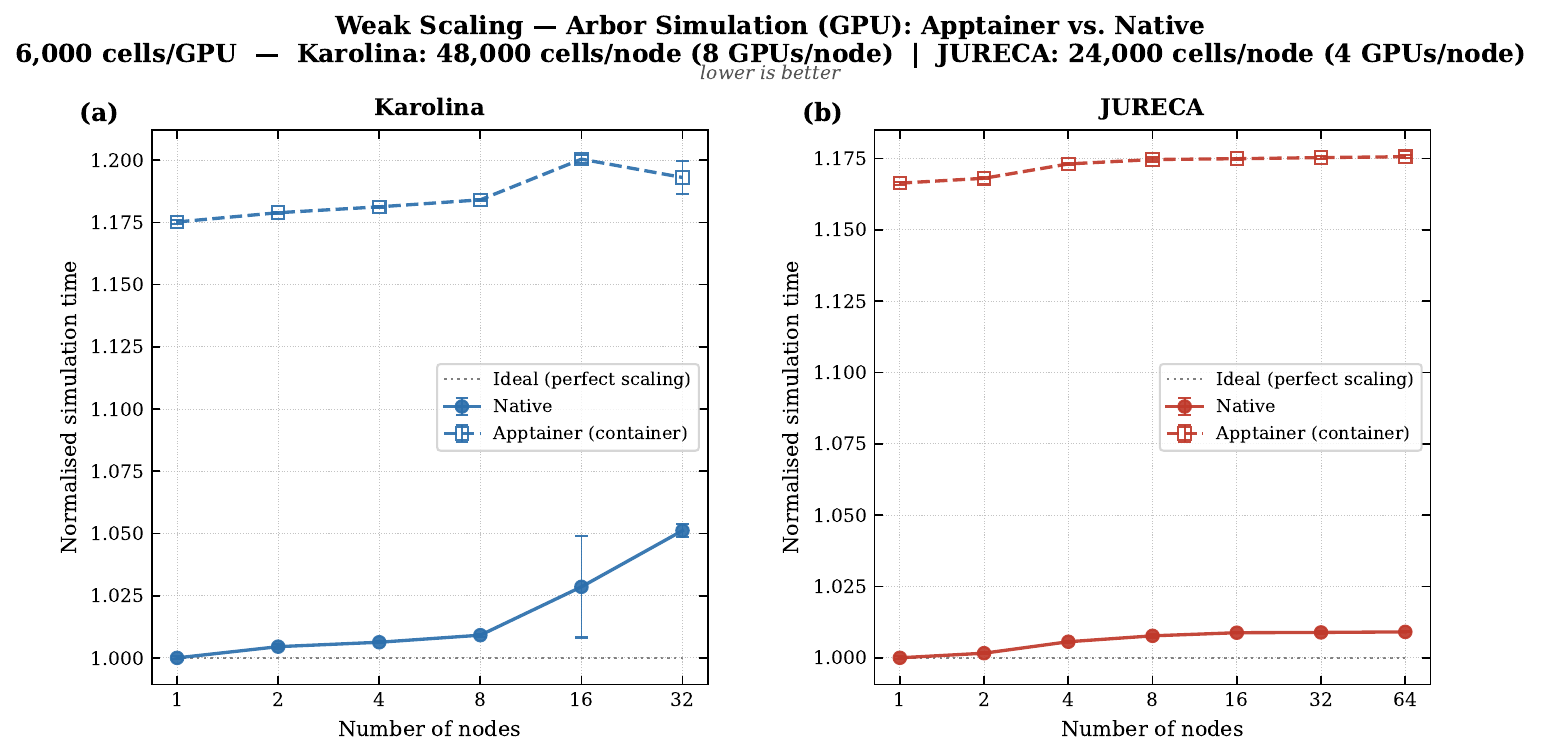}
	\caption{\normalfont Weak scaling performance of the Arbor ring network benchmark
			(GPU, \num{6000} cells per GPU; \num{48000} cells per node on Karolina
			with 8~GPUs per node; \num{24000} cells per node on JURECA with
			4~GPUs per node) on (a)~Karolina and (b)~JURECA.
			Simulation time is normalised to the single-node native baseline; a value of \num{1.0} (dotted line) indicates ideal scaling. The container introduces a constant offset of approximately \SI{17}{\percent} above the native baseline that does not grow with node count,
			confirming that Apptainer does not impair GPU scaling behavior. Error bars denote the standard deviation across repeated runs. Lower is better.}
	\label{fig:ArborGpuWS}
\end{figure}

\paragraph{Weak Scaling (GPU)}  The GPU weak scaling results (\num{6000} cells per GPU; \num{48000} cells per node on Karolina with 8~GPUs per node; \num{24000} cells per node on JURECA with 4~GPUs per node) are presented in \cref{fig:ArborGpuWS}.
The results demonstrate that Apptainer again introduces an overhead relative to native execution, while both environments scale near-ideally as the problem size grows.

On Karolina (\cref{fig:ArborGpuWS}a), the native simulation completes in \SI{78.1}{\second} at a single node and remains within \SIrange{78.1}{82.1}{\second} ($+\SI{5.1}{\percent}$) across all node counts, demonstrating near-ideal weak scaling. The container---starting at \SI{91.8}{\second} at 1 node---is offset by a stable $+\SIrange{13.4}{13.7}{\second}$ ($+\SIrange{16.7}{17.5}{\percent}$) above the native baseline for 1--16 nodes; this difference is entirely attributable to the $+\SI{17.5}{\percent}$ gap already present at 1 node and does not grow with scale, confirming that the PMIx-based inter-node communication path inside the container introduces no additional overhead as the job size increases.

On JURECA (\cref{fig:ArborGpuWS}b), the native simulation completes in \SI{75.0}{\second} at a single node and remains within \SIrange{75.0}{75.7}{\second} ($+\SI{0.9}{\percent}$) across the full 1--64 node range, demonstrating near-ideal weak scaling. The container---starting at \SI{87.5}{\second} at 1 node---maintains a remarkably constant absolute offset 
of $+\SI{12.5}{\second}$ ($+\SI{16.6}{\percent}$) throughout, with no measurable degradation as the node count increases sevenfold.

The GPU container overhead is a consistent $\SIrange{12}{19}{\percent}$ slowdown in simulation time (${\sim}\SI{13}{\second}$ in absolute terms for weak scaling). This overhead does not constitute a scaling penalty: under strong scaling,  where the total workload is fixed and distributed across an increasing  number of GPUs, the absolute difference decreases proportionally from 
${\sim}\SI{66}{\second}$ (${\sim}\SI{18}{\percent}$) at 1 node to ${\sim}\SI{1}{\second}$ (${\sim}\SI{13}{\percent}$) at 64 nodes, consistent with a constant relative overhead applied to a shrinking per-GPU workload. Under weak scaling, where the per-node workload remains constant, both the absolute penalty (${\sim}\SI{13}{\second}$) 
and the relative overhead (${\sim}\SI{16.6}{\percent}$) are invariant across all node counts, further confirming that the overhead scales  with the computational load rather than the number of processes. The origin of this overhead is not 
yet fully understood; kernel-level benchmarking would be required to isolate the precise source, and we leave this to future investigation. Given that the same image ran without modification on two distinct HPC clusters, this overhead may represent an acceptable trade-off between portability and raw performance for production neuroscience workloads.

\section{Discussion}

\paragraph{Initialization vs.\ Runtime Overhead:}
As noted in our limitations, the GPU Arbor container 
introduces a constant relative overhead of $\SIrange{12}{19}{\percent}$ in simulation time.
Interestingly, this overhead is absent in other CPU runs (for both Arbor and NEURON) and is not reflected in our NCCL inter-GPU communication microbenchmark runs. However, we did not evaluate the influence of different CUDA versions between host and containerized environment (\cref{tab:software}).

While the containerization introduces I/O overhead stemming from the additional base layer from the image, the number of I/O operations on network file systems is minimized as all software is packaged into one image file.
Consequently, the exact mechanism behind this GPU-specific initialization delay in Arbor runs remains unclear and will be a focus of future investigation.

\paragraph{Feasibility of Portable \gls{hpc} Containers:}
Our evaluation confirms the feasibility of containerizing complex, MPI- and CUDA-enabled \gls{hpc} tools without sacrificing execution speed.
Deploying our representative Apptainer images across diverse systems (e.g., Karolina, JURECA) yielded negligible communication overhead in microbenchmarks and maintained near-bare-metal parallel efficiency for CPU-bound simulators.
While GPU-accelerated workloads exhibited a minor overhead overall computational scaling remained robust, validating that portability and performance can coexist in distributed environments.

\section{Outlook}

\paragraph{Paving the Way for Automated \gls{esd} Integration:}
While the \gls{esd} already successfully provides an automated, \gls{ci}-driven software environment for native \gls{hpc} deployments and standard non-\gls{hpc} containers, these findings serve as the foundation for our next major objective:
integrating portable \gls{hpc} container generation directly into the \gls{esd}'s \gls{ci} pipeline.
Currently, domain scientists face high barriers and insufficient support for self-deploying complex toolchains involving a broad spectrum of tools natively on \gls{hpc} systems.
By offloading the compilation and validation burden to the \gls{esd}'s automated \gls{ci} process, we will enable researchers to easily pull pre-compiled, performance-verified images, drastically lowering the barrier to entry.

\paragraph{Containerized Benchmarks as a Diagnostic Tool for HPC System Verification:}
The benchmarking approach proved effective as an automated infrastructure verification tool, extending beyond simple portability assessment. By systematically comparing native performance against a containerized baseline, we uncovered hidden production system misconfigurations in both MPI startup latency (\texttt{osu\_init}) and collective communication bandwidth (via \texttt{nccl-test}) on JURECA-DC that had gone undetected through conventional monitoring. The containerized environment served as a controlled, reproducible reference, making native regressions immediately quantifiable; after reporting these findings to the system administrators, corrective changes were identified and implemented.
This demonstrates that a rigorous dual-environment benchmarking strategy provides not only a portability guarantee but also a reliable, reproducible standard for validating correct software stack deployment and ensuring the absence of performance regressions.

\paragraph{Future Host-System Optimization:}
To maximize the efficiency of these future \gls{esd} containers, our next step is to improve automated host-system integration.
We plan to automate the utilization of site-installed system packages --- for example, by automatically detecting and registering global and site-local EESSI-deployed software packages into our Spack-based build process.
This will allow the containerized environment to seamlessly and automatically leverage pre-optimized host interconnect drivers and libraries without manual user intervention.
Initial investigations \cite{mueller2025ebrains} used a transparent base image approach, where only the \gls{esd}-specific components were included in the image and the system dependencies were transparently bind-mounted into the container.

\paragraph{Community Trust and Data-Sensitive Subsets:}
Ultimately, a transparent, performance-verified deployment flow fosters community trust by guaranteeing reproducible science.
Furthermore, transitioning the \gls{esd} to a containerized model facilitates the creation of verified environment ``subsets''.
This is particularly critical for clinical neuroscience;
researchers will be able to deploy minimal, secure container subsets into isolated environments to process sensitive patient data, knowing the computational behavior remains strictly identical to the public \gls{esd} release.

\paragraph{Streaming Software:}
Much like EESSI deployments \parencite{droge2023eessi}, our approach described in \cref{sssec:sw_builds} will enable on-demand ``software streaming''.
However, in contrast to a fully expanded installation, we expect differences in terms of startup latency:
our solution's access granularity is coarser than flat installations provided via CVMFS, but the overall number of non-local file system accesses is significantly reduced.
Regardless, our approach enables users to efficiently compose use-case-specific containers that execute natively on non-EESSI-enabled systems.
A comprehensive performance evaluation of this streaming mechanism, specifically quantifying the impact of these I/O trade-offs on application initialization at scale, will be a primary focus of our future benchmarking efforts.

\section*{Acknowledgements}
This work has received funding from
the EC Horizon 2020 Framework Programme
under grant agreement 
945539 (HBP SGA3), 
the EC Horizon Europe Framework Programme
under grant agreement
101147319 (EBRAINS 2.0).
We gratefully acknowledge the EuroHPC Joint Undertaking for awarding this project access to computing resources under benchmark grant EU-25-119.

\printbibliography

\end{document}